\documentclass[12pt,preprint]{aastex}

\usepackage{epsfig}

\shorttitle{C$_{60}$ in IRAS 01005+7910}
\shortauthors{Zhang \& Kwok}

\begin{document}

\title{ Detection of C$_{60}$ in the proto-planetary nebula IRAS 01005+7910
}

\author{Yong Zhang and Sun Kwok}
 \affil{Department of Physics, University of Hong Kong, Pokfulam, Hong Kong, China}
 \email{zhangy96@hku.hk; sunkwok@hku.hk}

\begin{abstract}

We report the first detection of buckminsterfullerene (C$_{60}$) in a proto-planetary nebula (PPN). 
The vibrational transitions of C$_{60}$ at 7.0, 17.4, and 18.9\,$\mu$m are detected in the
{\it Spitzer}/IRS spectrum of  IRAS 01005+7910.  This detection suggests that fullerenes are formed shortly after the asymptotic giant branch but before the planetary nebulae stage.
A comparison with the observations of C$_{60}$ in other sources is made and the implication on circumstellar chemistry is discussed.

\end{abstract}

\keywords{infrared: stars ---
stars: AGB and post-AGB --- stars: circumstellar matter
}

\section{Introduction}

Fullerenes, together with other forms of carbon such as graphite, diamonds, and carbynes, are expected to be a important components of the interstellar medium \citep{hen98}.
The most stable fullerene is buckminsterfullerene (C$_{60}$) which has a soccer-ball like structure  \citep{kroto}.
Fullerenes have been proposed as possible carriers of diffuse interstellar bands \citep[e.g.][]{kroto,leger}, the origin of which is a long-standing mystery \citep[see][for a review]{herbig95}. \citet{foing94} found that the laboratory spectrum of  C$_{60}^+$  in argon and neon matrices shows approximate matches with two diffuse interstellar bands in the near infrared spectra of stars. 
Possible formation processes of fullerenes in space that have previously been discussed include condensation in supernova gas, shock-induced decomposition of hydrogenated amorphous carbon grains, cold interstellar gas-phase chemistry, etc \citep[see][and references there in]{moutou99,sel10}.  The carbon-rich hydrogen-poor circumstellar envelopes, such as Wolf-Rayet (WR) stars and R Coronae Borealis stars, have also been proposed as favorable sites for the synthesis of fullerenes \citep{che00,goeres92}.  However, none of these theoretical predictions have been confirmed observationally.

The search for fullerenes in space started soon after their synthesis in the laboratory.  The early efforts  to search for the electronic transitions of C$_{60}$ in the optical and UV wavelengths have resulted in no definite evidence showing its presence in space \citep[e.g.][]{snow89,herbig00,sassara01}. 
Another approach is to search for the infrared vibrational modes of the molecule \citep{clayton95}.
\citet{kwok99} noted a faint feature at 17.85 and 18.90\,$\mu$m in the {\it ISO}-SWS spectrum of the proto-planetary nebula (PPN) IRAS 07134+1005 and suggested that they could correspond to the $\nu_{27}$ and $\nu_{28}$ vibrational modes of the C$_{60}$ molecule.   However, these features are not confirmed by
subsequent {\it Spitzer} observations.  
A search of these bands in the reflection nebula NGC 7023 with {\it ISO} was also not successful \citep{moutou99}.
Only very recently, C$_{60}$ was unambiguously detected 
in the planetary nebula (PN) Tc~1  \citep{cami10} and the reflection nebulae NGC 7023 and NGC 2023 \citep{sel10}. This was followed by the detections of C$_{60}$ in four more PNs \citep{gar10}. 

These detections of C$_{60}$ in objects at the late stages of stellar evolution raise the concrete question of how C$_{60}$ is formed in circumstellar environment.  Since the element C is synthesized in Asymptotic Giant Branch (AGB) stars and many C-based molecules have been detected in the outflow of AGB stars, it is reasonable to expect the C$_{60}$ can also be one of the products of circumstellar molecular synthesis.
Laboratory results show that the formation efficiency of fullerenes depends on the content of hydrogen and this has led to the suggestion by \citet{cami10} that under hydrogen-poor environment, the formation of C$_{60}$ is favored, and otherwise the chemical pathway favors the formation of Polycyclic Aromatic Hydrocarbon (PAH) molecules. This scenario is supported by  the fact that the infrared spectrum of Tc~1 does not exhibit the aromatic infrared bands (AIBs) commonly assigned to as arising from PAH molecules.  
However, contrary to this scenario, C$_{60}$ was detected in hydrogen-containing PNs showing the AIBs  \citep{gar10}. The detection of C$_{60}$ in NGC~7023   \citep{sel10}, an object with strong AIB emissions, also confirms that C$_{60}$ and the AIB carriers can coexist.
As an alternative explanation, 
\citet{gar10} suggested that both PAH molecules and fullerenes are formed from the destruction of hydrogenated amorphous carbon (HAC), and the non-detection of AIBs in Tc 1 is due to the longer survival time of C$_{60}$ molecules. 
A search for  C$_{60}$ in a large sample of R Coronae Borealis (R Cor Bor) stars was performed by \citet{gar11}. C$_{60}$ is clearly detected in one of the R Cor Bor stars showing AIBs, but C$_{60}$ is absent in the most hydrogen-poor ones.  These results further support the premise that fullerene and the carrier of the AIBs can coexist.


PPNs are the descendants of the asymptotic giant branch (AGB)  stars and the immediate precursors of PNs \citep{kwok93}.   The changes of infrared spectroscopic properties in AGB stars, PPNs, and PNs have been known as a consequence of evolution \citep{kwok04}.  Thus PPNs can provide a unique opportunity to investigate the formation history of C$_{60}$ in circumstellar envelopes. We have obtained infrared spectra of a sample of PPNs
\citep{zhang10} which cover the wavelengths of four C$_{60}$ vibrational transitions at 7.0 ($\nu_{25}$), 8.5 ($\nu_{26}$), 17.4 ($\nu_{27}$), and 18.9\,$\mu$m ($\nu_{28}$) \citep{nem94}.  These spectra therefore can serve as a platform to search for C$_{60}$ in PPNs.

In this paper we report a new detection of C$_{60}$ in a PPN, IRAS 01005+7910. Unlike the other sources in our sample, IRAS 01005+7910 does not show the 21-$\mu$m feature and its central star has a higher temperature ($\sim21,500$\,K), suggesting that it a PPN about to enter the PN stage \citep{zhang10}.
\citet{hu02} classified it as a B2 Ie star with V magnitude of 10.85. Its hydrogen Balmer lines show P Cygni profiles. Through a study of the high-resolution spectrum, \citet{klo02} concluded that it is a carbon-rich PPN with a luminosity $\log(L/L_\sun)=3.6$ at a distance about 3\,kpc.

\section{Data}

The study makes use of the infrared spectra retrieved from the 
{\it Spitzer Heritage Archive} (SHA) \footnote{http://sha.ipac.caltech.edu/applications/Spitzer/SHA/}. The observations were conducted with the Infrared Spectrograph (IRS; Houck et al. 2004) on the {\it Spitzer Space Telescope} \citep[{\it Spitzer};][]{wer04} in 2004 and 2006. We have carried out a systematic search for C$_{60}$ in the ten PPNs studied by \citet{zhang10} and found that among the studied sample IRAS 01005+7910 is the unique one that clearly exhibits the C$_{60}$ features.  The spectra of IRAS 01005+7910 was obtained with the short-wavelength low-resolution module (SL; 5--14.5\,$\mu$m) as part of the program 30036 (PI: G. Fazio), and the short-wavelength high-resolution module (SH; 9.5--19.5$\mu$m) as part of the program 93 (PI: D. Cruikshank). Details of the data processing have been described elsewhere  \citep[e.g.][]{hri09,cer09} and are not repeated here.
However, no observation utilizing long-wavelength high-resolution module (LH; 18.7--37.2\,$\mu$m) was made for IRAS 01005+7910.


\section{Results}

\begin{figure*}
\epsfig{file=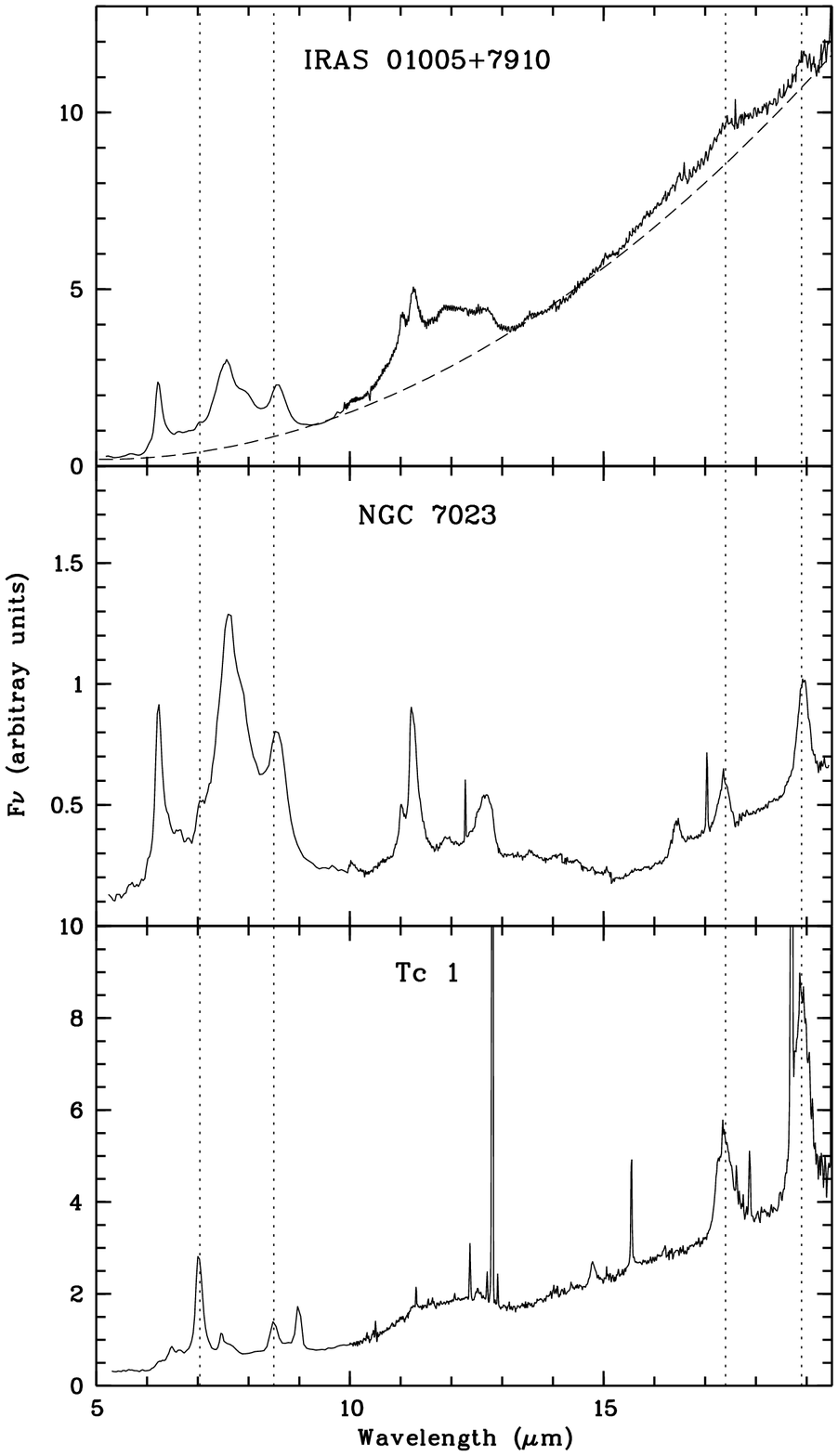, height=21cm}
\caption{The {\it Spitzer}/IRS  spectrum of IRAS 01005+7910 compared to the IRS spectra of two other known C$_{60}$ sources (NGC~7023 from Sellgren et al. 2010, and Tc~1 from Cami et al. 2010). The vertical dotted  lines mark the wavelengths of the C$_{60}$ lines at 7.0, 8.5, 17.4, and 18.9\,$\mu$m. The dashed line represents a fit to the continuum.  The other prominent features at 6.2, 7.7, 8.6, 11.3, and 12.7 $\mu$m are AIBs.}
\label{spe}
\end{figure*}

Figure~\ref{spe} shows that the {\it Spitzer}/IRS spectrum  of IRAS 01005+7910  is dominated by a thermal dust continuum, the 11.5\,$\mu$m SiC emission,
the aromatic infrared bands (AIBs) at 6.2, 7.7--7.9, 8.6, 11.3, and 12.7\,$\mu$m, and the
15--20\,$\mu$m plateau feature. 
The AIB  and the broad plateau emission features have been previously detected and discussed by \citet{cer09} and \citet{zhang10}.  In this paper, we report the detection of the  C$_{60}$ features at $7.04\pm0.05$,  $17.4\pm0.05$, and $18.9\pm0.04$\,$\mu$m.  The fourth expected C$_{60}$ feature at 8.5\,$\mu$m feature is badly blended with the AIB 8.6\,$\mu$m feature. These C$_{60}$ features have a width of
$0.31\pm0.05$\,$\mu$m, much broader than the spectral resolution.  
The measured widths of the C$_{60}$ features are similar to those seen in other PNs \citep[see Table 1,][]{gar10}.
These C$_{60}$ features are not found in any other PPNs in our sample.  This suggests that these features likely share a common origin, and thus strengthens the identification of C$_{60}$ as their carrier.

\begin{figure*}
\epsfig{file=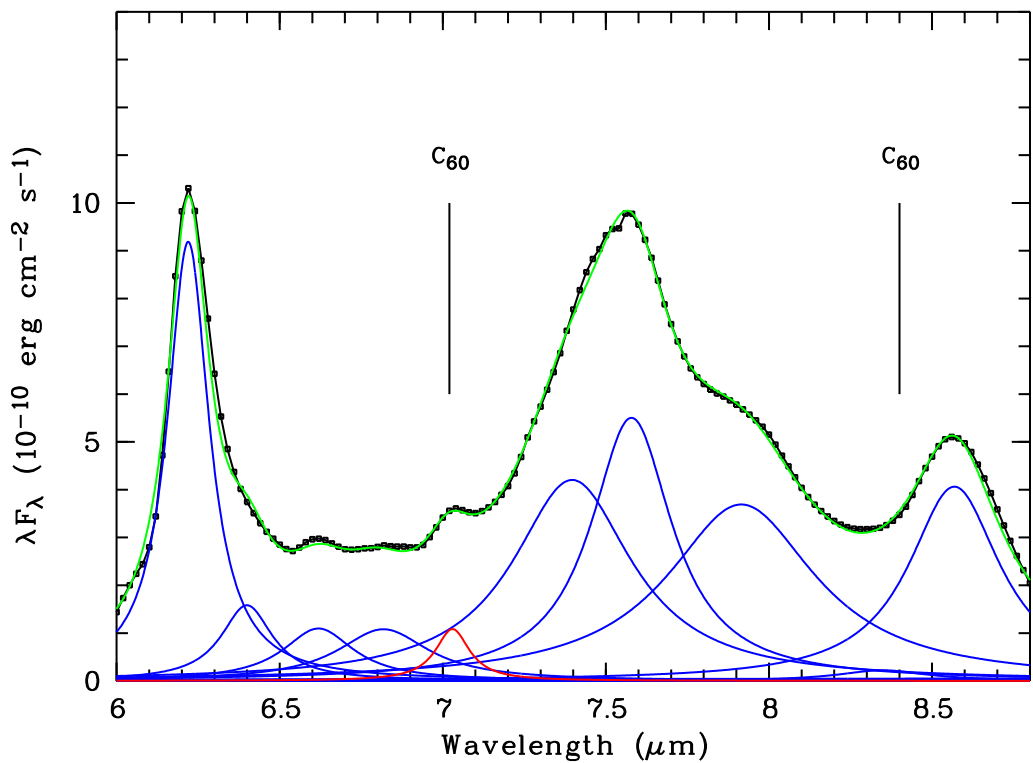, height=10cm}
\caption{Continuum-subtracted {\it Spitzer}/IRS  spectrum of IRAS 01005+7910 (black dots) between 6 and 8.8 $\mu$m.  The expected positions of the C$_{60}$ features at 7.04 and 8.45\,$\mu$m are marked.  The blue curves show the model fits to the individual AIBs (a list of the features is given in the text), with the total model spectrum given as the green curve.  The red curve is a model fit to  the 7.0\,$\mu$m  C$_{60}$ feature. The 8.5\,$\mu$m  C$_{60}$ line is badly blended with the 8.6\,$\mu$m AIB feature and is not detected.}
\label{c60a}
\end{figure*}

\begin{figure*}
\epsfig{file=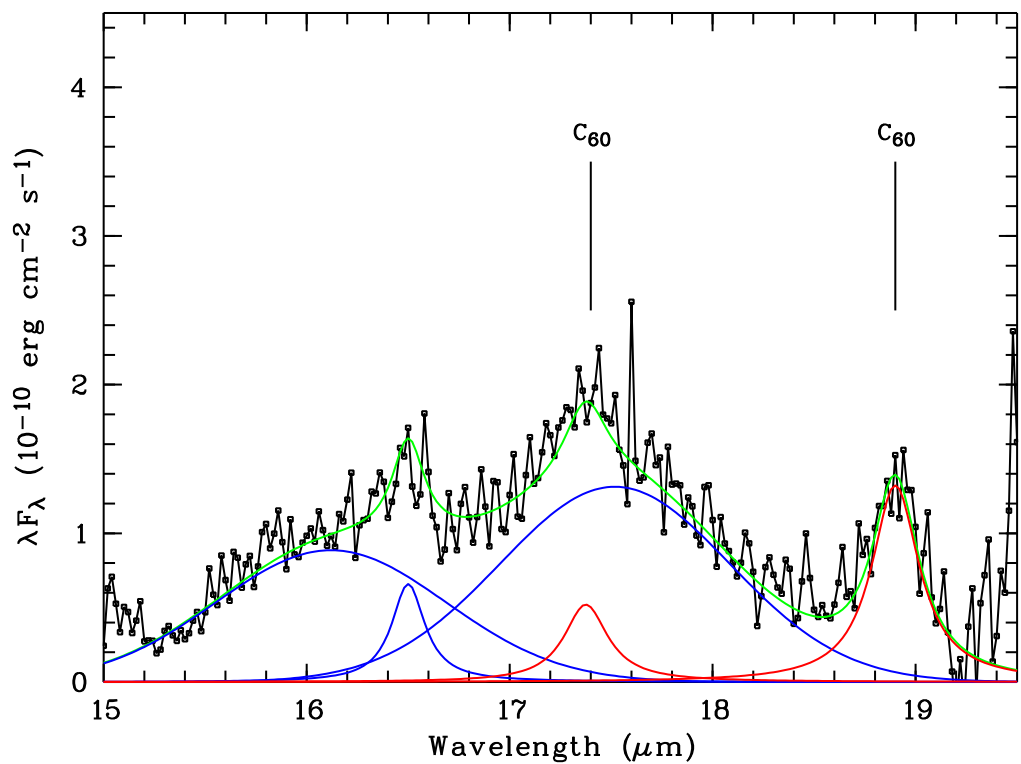, height=10cm}
\caption{Continuum-subtracted {\it Spitzer}/IRS spectrum of IRAS 01005+7910 (black curve) between 15 and 19.5 $\mu$m.  The positions of the C$_{60}$ lines at 17.4 and 18.9\,$\mu$m are marked. The blue curves show the model fits to the individual dust/plateau features, the red curves are the model fits to the 17.4 and 18.9\,$\mu$m  C$_{60}$ lines, and the total model spectrum is given by the green line.
 Note that the 15--20\,$\mu$m plateau is fitted  by two Gaussian profiles.}
 
\label{c60b}
\end{figure*}

The continuum was fitted using the feature-free spectral regions, and was subtracted from the observed spectrum. 
As the features in PPN spectra are usually broad and blended with each other, we conducted a spectral decomposition using the IDL package PAHFIT developed by \citet{smi07} in order to accurately measure the feature fluxes. Drude profiles are assumed for the AIBs and C$_{60}$  features. The actual profile of  the 15--20\,$\mu$m plateau, where C$_{60}$ 17.4 and 18.9\,$\mu$m features are superimposed, is poorly known. We have assumed two broad Gaussian profiles having a width of 1.3\,$\mu$m and peaked at 16.1 and 17.5\,$\mu$m for the plateau (this causes only slight uncertainty in the flux measurements since the  C$_{60}$ features are much narrower than the plateau).  For the fitting, we have taken into account the AIBs at 6.2, 6.4, 6.6, 6.8, 7.4, 7.6, 7.9, 8.3, 8.6, and 16.5 $\mu$m.
Figures~\ref{c60a} and \ref{c60b} give the zoomed-in view of these C$_{60}$ features and the fitting results.  From the fitting results, we derived  fluxes of  $(3.0\pm0.3)\times10^{-15}$\,W/m$^2$, $(1.2\pm0.5)\times10^{-15}$\,W/m$^2$, and $(2.9\pm0.5)\times10^{-15}$\,W/m$^2$ for the C$_{60}$ 7.0, 17.4, and 18.9\,$\mu$m features, respectively.  The errors of the fluxes were estimated using the full covariance matrix of the least-squares parameters.
\citet{sel10} found that the C$_{60}$ $17.4$\,$\mu$m transition in the spectrum of NGC 7023 is partially blended with an AIB feature. If this is also the case for IRAS 01005+7910, the estimated strength of the $17.4$\,$\mu$m feature would be the upper limit.


\citet{hri00} presented the {\it ISO}-SWS spectrum of IRAS 01005+7910 covering a wavelength range of 2.4--45.4\,$\mu$m.  Due to the lower sensitivity of the {\it ISO} SWS compared to the IRS, the C$_{60}$ features are completely overwhelmed by the noise in the {\it ISO} spectrum.   However, a strong feature at 30\,$\mu$m is detected by {\it ISO}, and its presence was subsequently confirmed by the {\it Spitzer}/IRS spectrum \citep{cer09,zhang10}.
This seems to support the finding by \citet{gar10} that all the C$_{60}$ sources exhibit
the 30\,$\mu$m feature. However, this correlation only applies to circumstellar sources as the {\it Spitzer} archive spectra of the reflection nebulae NGC 2023 and NGC 7023 (program 40276, PI: K. Sellgren) do not show the 30\,$\mu$m feature.
The 30\,$\mu$m feature is commonly seen in carbon-rich AGB stars, PPNs, and PNs  \citep{for81,vol02}, and has been attributed to solid magnesium sulfide \citep[MgS;][]{goe85}.  However, the identification of MgS as the carrier of the 30\,$\mu$m feature  is debatable as this feature is only detected in carbon-rich sources. Recently, \citet{zha09} found that the MgS dust mass in circumstellar envelopes is not enough to account for the observed feature strength. Therefore, carbonaceous compounds might be more likely to be the carrier of the 30\,$\mu$m feature.   In order to establish a connection between the 30\,$\mu$m feature with the C$_{60}$ features, more C$_{60}$  sources need to be discovered.

In Figure~\ref{spe}, we compare the {\it Spitzer}/IRS spectra of IRAS 01005+7910 with two other C$_{60}$ sources, NGC 7023 \citep{sel10} and Tc 1 \citep{cami10}. 
All three sources have a strong infrared excess and IRAS 01005+7910 and Tc 1 have a very red (low color temperature) continuum.  
IRAS 01005+7910 and NGC 7023 show strong AIB features, which are absent in Tc 1. The fact that IRAS 01005+7910 does not show the narrow atomic lines as seen in the spectrum of Tc 1 is consistent with the object being a PPN and its central star is not hot enough to ionize the surround envelope.  After subtracting the continuum, we found that the spectral shape of IRAS 01005+7910  is similar to that of NGC 7023. \citet{cami10} also detected a few
weaker C$_{70}$ features in Tc 1, which are not seen in IRAS 01005+7910 and NGC 7023.  
Assuming that all the sources have the same C$_{70}$/C$_{60}$ strength ratio, the C$_{70}$ features in IRAS 01005+7910 should be well below the detection limit.

\section{Discussion}

\begin{figure*}
\epsfig{file=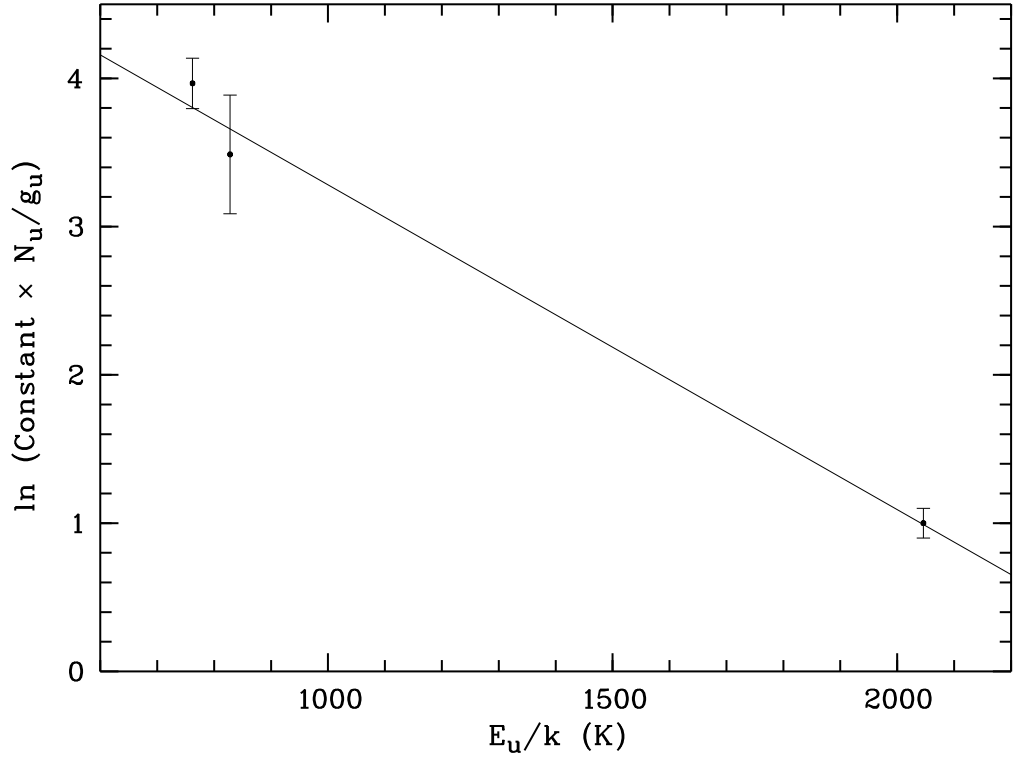, height=10cm}
\caption{Vibration diagram for the detected C$_{60}$ bands in IRAS 01005+7910.  
$N_u$, $g_u$, and $E_u$ are the population, statistical weight, and excitation energy of the upper level, respectively.
}
\label{ro}
\end{figure*}

The relative intensities of the C$_{60}$ lines reflect the excitation conditions of the molecule.
Assuming a thermal distribution of the vibrational states, \citet{cami10} determined the excitation temperature of Tc~1 to be $\sim330$\,K, and suggested that the C$_{60}$ molecules are in a solid.  
Similar temperature values were also derived in the PNs studied by \citet{gar10}.
Using a similar procedure as   \citet{cami10} and \citet{gar10}, we have constructed a vibrational diagram for IRAS 01005+7910 from the observed fluxes of the three C$_{60}$ lines (Figure~\ref{ro}).  An excitation temperature of $460\pm50$\,K is derived.

The C$_{60}$ line ratios in IRAS 01005+7910 are $I_{7.0}/I_{18.9}=1.0\pm0.3$ and $I_{17.4}/I_{18.9}=0.4\pm0.2$, which are very close to the values found by in NGC 7023 \citep[$I_{7.0}/I_{18.9}=0.82\pm0.12$ and $I_{17.4}/I_{18.9}=0.42\pm0.02$;][]{sel10}.  Since the central star of IRAS 01005+7910 has a similar temperature as NGC 7023 ($\sim17,000$\,K), it is possible that the C$_{60}$ molecules in IRAS 01005+7910 and NGC 7023 are excited in a similar manner.  \citet{sel10} suggests that 
C$_{60}$ molecules in NGC 7023 are in the gas phase and they are excited by UV photons from the central star followed by radiative cascade.
If this is also the case for IRAS 01005+7910, then the excitation temperature derived above is just a representation of population distribution and is not related to other physical quantities such as kinetic temperature of the gas.

Laboratory measurements show that the wavelengths of gas-phase C$_{60}$ bands shift with temperature
\citep{frum91,nem94}. The positions of the four infrared bands $\nu_{25}$, $\nu_{26}$, $\nu_{27}$, and $\nu_{28}$ shifts from 6.97, 8.40, 17.41, and 18.82 at 0\,K to 7.11, 8.55, 17.53, 18.97 at 1000\,K, respectively.  Our measurements show that the wavelengths of all C$_{60}$ bands in IRAS 01005+7910 lie inside these ranges.  Assuming that the frequencies of the C$_{60}$ bands have a linear dependence on the temperature, we can estimate the temperature as 100--600\,K.
The observed widths of the C$_{60}$ features are about 0.3 $\mu$m (see Section 3), which correspond to a wavenumber width of $\sim$60 cm$^{-1}$ and $\sim$10 cm$^{-1}$ for the 7 and 17.4/18.9 $\mu$m bands respectively.  These values can be  compared with the laboratory measured widths of about $\sim$13 cm$^{-1}$ of gas-phase C$_{60}$ \citep{frum91}, suggesting that  a gas-phase origin of the molecule cannot be ruled out.




We estimate the abundance of C$_{60}$ following the same method of \citet{sel10} by calculating the total strength ratios between the C$_{60}$ and AIB features and assuming that the carrier of the AIB features to contain $6\pm2$\% of interstellar carbon \citep{cer09}. The strength ratio of C$_{60}$ to AIB emission is 0.01 in the observed wavelength range of IRAS 01005+7910, resulting in a percentage of carbon in C$_{60}$  of $0.06\pm0.02$$\%$. This value is slightly lower than those obtained in NGC 7023 \citep{sel10} and the PN SMP SMC 16 \citep{gar10}, but a factor of 25 lower than that of Tc~1 estimated
by \citet{cami10}.

With the detection of C$_{60}$ in circumstellar envelopes, the next question is how they are formed.  
In the laboratory, C$_{60}$ can be effectively produced from the vaporization of graphite in a hydrogen-poor environment. \citet{cami10} proposed that fullerenes are produced only in hydrogen-poor envelopes created by a late AGB thermal pulse.   However, \citet{gar10} argued that C$_{60}$ can be synthesized under  hydrogen-containing environment. As shown in Figure.~\ref{spe}, the spectrum of IRAS 01005+7910 exhibits both hydrogen-containing AIBs and C$_{60}$ features, supporting the latter hypothesis.  \citet{gar10} also suggested that both C$_{60}$ and PAHs are the products of decomposition of HACs. In this scenario, photochemical processing can lead to dehydrogenation of the dust grains and form PAHs and  C$_{60}$ molecules \citep{scott97}. On the other hand, the dehydrogenation of dust grains can also induce the formation of H$_2$ in the grain surfaces \citep[e.g.][]{fleming10}. The strongest H$_2$ line in the observed wavelength range is the 0--0 S(0) transition at 28\,$\mu$m. This line is not detected in the spectrum of IRAS 01005+7910 \citep{hri00,cer09,zhang10}. Moreover, we have detected H$_2$ in two PPNs with no detectable C$_{60}$  \citep{zhang10}. Because of this lack of correlation between the presence of H$_2$ and C$_{60}$, we are unable to give addition support for the idea that the formation of C$_{60}$ is the result of dehydrogenation of HACs.

Our detection suggests that fullerenes can be formed in the PPN stage. So far, there is no definite detection of C$_{60}$ in AGB stars.  Although \citet{clayton95} noted a possible emission feature centered at 8.6\,$\mu$m in the spectrum of the bright AGB star IRC+10216, the other  C$_{60}$ features were not detected in the {\it ISO}-SWS spectrum \citep{cer99}.
The circumstellar spectra of AGB stars are dominated by silicates or silicon carbide emission features \citep{kwok1997}. 
Although there are a small number of  AGB stars exhibiting AIBs,
the AIBs mainly emerge in the post-AGB phase. 
Is it possible that the formation of C$_{60}$ is related to the emergence of the AIB features?
\citet{sel10} found the C$_{60}$ and AIB emissions in NGC 7023 have different spatial distributions and attributed this to the effect of UV-excitation.  It can be argued that the C$_{60}$ and the AIB carriers are already present in the AGB phase of evolution but are not excited until the stars evolve to the PPN phase.  However, comparisons between the spectra of AGB stars, PPNs, and PNs suggest a sequence of molecular synthesis, with acetylenes forming in the late AGB phase, leading to the formation of diacetylenes, triacetylenes, and  benzene in the PPN phase \citep{kwok04}.  Since these molecules are detected in absorption, the question of excitation does not arise.  Since benzene is the first step toward the synthesis of aromatic materials, we can say with confidence that aromatic compounds only form after the AGB.  If C$_{60}$ molecules are synthesized during the AGB, they should be detectable with absorption spectroscopy.


\section{Conclusions}

The detection of C$_{60}$ in a PPN as reported in this paper, together with the detection of this molecule in 5 PNs, confirm that the late stages of stellar evolution is a phase of active molecular synthesis.  Beginning with simple diatomic molecules such as CO, CN, C$_2$, dozens of gas-phase organic molecules have been seen in the stellar winds from AGB stars.  The formation of acetylene  during the late AGB phase is believed to lead to the formation of benzene in the post-AGB phase of evolution. This also coincides with the first  detection of vibrational modes of aromatic and aliphatic compounds.  From this study, we now learn that   gas-phase C$_{60}$ molecules may also form during the same epoch.   
As the number of C$_{60}$ detection increases, we would be in better position to study the relationships between C$_{60}$ and the carriers of other spectral features such as the AIBs, and 21- and 30-$\mu$m features. 

The detection of C$_{60}$ in the outflows from evolved stars also raises the possibility of the molecule being detected in the diffuse interstellar medium as the molecule is stable and should be able to survive journeys through the interstellar medium \citep{foing94}.  The fact that a large variety of presolar grains (e.g., SiC) have been detected in meteorites \citep{zinner1998} raises the possibility that presolar    C$_{60}$ can be incorporated into comets and asteroids and be detected in meteorites.   In fact,  
C$_{60}$ and C$_{70}$, as well as higher fullerenes, have been detected in the Allende meteorite \citep{becker99}.
The evidence for stellar synthesis of C$_{60}$ in the late stages of stellar evolution as presented in this paper therefore adds further support to the idea of chemical enrichment of the Solar System by stellar molecular products.

\acknowledgments

We are grateful to Peter Bernath for helpful discussions. We also thank 
 Kris Sellgren and Jan Cami for providing the IRS spectra of NGC~7023 and
Tc~1 from the their papers (Sellgren et al. 2010, and Cami et al. 2010).
This work is based on observations made with the {\it Spitzer Space Telescope}, which is operated by the Jet Propulsion Laboratory, California Institute of Technology, under a contract with NASA.
This work was supported by a grant to SK from the Research Grants Council of the Hong Kong Special Administrative Region, China (Project No. HKU 7020/08P) and a grant to YZ from the Seed Funding Programme for Basic Research in HKU (200909159007).

\clearpage


\end{document}